\documentclass[runningheads]{llncs}
\usepackage[T1]{fontenc}
\usepackage{graphicx}

\usepackage[pdftex,dvipsnames]{xcolor}
\usepackage[pdftex,
            pdfauthor={R. Bögli et al.},
            pdftitle={A Systematic Literature Review on a Decade of Industrial TLA+ Practice (Preprint)},
            pdfsubject={Preprint -- Published in 19th Int. Conference on Integrated Formal Methods 2024},
            pdfkeywords={Formal Methods, TLA+, PlusCal, Literature Review, Industry}]{hyperref}
\hypersetup{
    pdfpagemode={UseOutlines},
    bookmarksopen=true,
    bookmarksopenlevel=1,
    bookmarksnumbered=true,
    bookmarksdepth=3,
    hypertexnames=false,
    colorlinks=true,
    linkcolor=Black,
    citecolor=ForestGreen,
    urlcolor=blue,
    pdfstartview={FitV},
    unicode=true,
    breaklinks=true
}

\usepackage{xurl}

\usepackage{multibib}
\newcites{ReferencesSectionFirst,ReferencesSectionSecond}{References (The 16 High-Affinity Papers),References (Other)}

\usepackage{bbding}
\newcommand{\orcid}[1]{\href{https://orcid.org/#1}{\includegraphics[width=10pt]{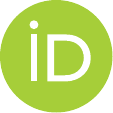}}}

\usepackage{xargs}
\newcommand{\TLA}{TLA\textsuperscript{+}\kern0.19em}
\newcommand{\TLAno}{TLA\textsuperscript{+}\kern-0.12em}

\usepackage{float}
\usepackage{caption}
\newcounter{listing}
\newcommand{\listingcaption}[1]{\refstepcounter{listing}\par\medskip\centering\textbf{Listing \thelisting.} #1\par\medskip}
\newcommand{\figurecaption}[1]{\refstepcounter{figure}\par\medskip\centering\textbf{Figure \thefigure.} #1\par\medskip}

\usepackage{enumitem}
\newcommand{\labelwithcolon}[1]{#1:}
\usepackage{tcolorbox}
\definecolor{custompurple}{RGB}{100, 0, 149}
\tcbset{
    mybox/.style={
        colback=custompurple!15,  
        colframe=custompurple!95,
        boxrule=0.5pt,
        left=1mm,
        right=1mm,
        top=1mm,
        bottom=1mm,
        rounded corners
    }
}

\begin{document}

\title{A Systematic Literature Review on a Decade of Industrial \TLA Practice\thanks{
	\emph{Preprint} -- Published in 19th Int. Conference on Integrated Formal Methods 2024,\\
	\hspace*{1.6cm}\url{https://doi.org/10.1007/978-3-031-76554-4_2}.
	}}
\titlerunning{SLR on Industrial \TLA Usage}

\author{
	Roman Bögli\inst{1}\orcid{0009-0004-8745-7800}\ \Envelope
	\and
	Leandro Lerena\inst{1}
	\and\\
	Christos Tsigkanos\inst{1,2}\orcid{0000-0002-9493-3404}
	\and\\
	Timo Kehrer\inst{1}\orcid{0000-0002-2582-5557}
}
\authorrunning{R. Bögli et al.}

\institute{
	Institute of Computer Science, University of Bern, Switzerland\\
	\email{\{roman.boegli,christos.tsigkanos,timo.kehrer\}@unibe.ch}
	\and
	Department of Aerospace, University of Athens, Greece
}

\maketitle

\begin{abstract}
\TLA is a formal specification language used for designing, modeling, documenting, and verifying systems through model checking.
Despite significant interest from the research community, knowledge about usage of the \TLA ecosystem in practice remains scarce.
Industry reports suggest that software engineers could benefit from insights, innovations, and solutions to the practical challenges of \TLAno.
This paper explores this development by conducting a systematic literature review of \TLAno's industrial usage over the past decade.
We analyze the trend in industrial application, characterize its use, examine whether its promised benefits resonate with practitioners, and identify challenges that may hinder further adoption.

\keywords{
    \begin{tabular}[t]{@{}l}
        Formal Methods \and
        \TLA \and
        PlusCal \and
        Literature Review \and \\
        \makebox[0pt][l]{Industry}
    \end{tabular}
}
\end{abstract}

\section{Introduction} \label{intro}

Despite the potential of increasing dependability of software-intensive systems, formal methods are still scarcely adopted in industry~\citeReferencesSectionSecond{reid2020MakingFormalMethods}.
A prominent example is the formal specification language \TLA (\emph{Temporal Logic of Actions})~\citeReferencesSectionSecond{lamport2002specifying}, which was designed for modeling, specifying, and verifying a variety of systems.
It provides a high-level mathematical notation to describe system behavior over time using temporal logic~\citeReferencesSectionSecond{pnueli1977TemporalLogicPrograms} and set theory, making it particularly well-suited for distributed or concurrent systems.
\TLA specifications can be verified using the \emph{TLC model checker}, which systematically explores all possible states and highlights property violations using counterexamples.
As the syntax of \TLA closely relates to the underlying mathematical notation, it can be challenging for software engineers that are accustomed to imperative programming languages.
Thus, Lamport~\citeReferencesSectionSecond{lamport2009PlusCalAlgorithmLanguage} introduced \emph{PlusCal}, a \texttt{C}-like programming language that translates into \TLA specifications and therewith aims to bridge the gap between conventional programming practices and formal methods. Unless stated otherwise, we use \TLA to refer to both \TLA and PlusCal in this paper.

\enlargethispage{\baselineskip}
\enlargethispage{\baselineskip}

\TLA promises several benefits in software development.
First, it allows for precise specification and verification of system behaviors and properties, substantially reducing the likelihood of fundamental design flaws or subtle bugs that traditional testing methods might miss.
Examples of these notoriously \mbox{hard-to-detect} errors include, for instance, race conditions and deadlocks.
Second, \TLA promotes clear and unambiguous system documentation, preventing misinterpretations that can arise from informal specifications written in natural human language.
This eases the integration of new system features and streamlines staff onboarding.
Lastly, the similarity of PlusCal's syntax to conventional programming languages lowers the entry barrier and encourages the adoption of formal methods in mainstream software development.

From an academic perspective, \TLA is appreciated due to its promising benefits, which is why it is suggested to practitioners in industry.
However, little is known about the actual adoption of \TLA in real-world settings and whether its promises resonate with practitioners.
To address this gap, we conduct a \emph{Systematic Literature Review} (SLR), guided by the following research questions:

\begin{enumerate}[label=\labelwithcolon{\textbf{RQ\arabic*}}, leftmargin=1.60cm,rightmargin=0cm, ref=\textbf{RQ\arabic*}]
	\item \label{rq1} What is the trend in \TLA usage in industry over the past decade?
	\item \label{rq2} How are industrial applications of \TLA characterized?
	\item \label{rq3} Do the promised benefits of \TLA resonate in industry reports?
	\item \label{rq4} What challenges hinder the adoption of \TLA in industry?
\end{enumerate}

\section{Review Methodology}

In terms of our SLR, we followed the guidelines in~\citeReferencesSectionSecond{kitchenham2007GuidelinesPerformingSystematic,petersen2015GuidelinesConductingSystematic} for collecting, filtering, and analyzing existing literature.
Moreover, following the guidelines by Carrera-Rivera et al.~\citeReferencesSectionSecond{carrera-rivera2022HowtoConductSystematic}, we employed \emph{\href{https://parsif.al}{Parsifal}} to streamline the SLR\@.
This online tool aids in creating the review protocol, including formulating inclusion and exclusion criteria, a quality assessment checklist, and a data extraction form.
Note that we use the term \emph{paper} as a synonym for publication, resource, report, or document.

Figure~\ref{papercollectionprocess} visualizes the paper gathering and reduction process over the sampled time period.
Each bar's total length corresponds to the number of optimistically found papers in a given year.
The subfigure on the right subsumes the totals per phase, which are further explained in the rest of this section.
The entire SLR was conducted by two researchers independently and disagreements were discussed and eliminated among all four authors.

\subsection{Optimistic Search}

Since this SLR focuses on the use of \TLA in industry, we chose to rely exclusively on \emph{\href{https://scholar.google.com/schhp?hl=en&as_sdt=0,5}{Google Scholar}} search results.
This platform includes grey literature, such as industrial reports, which may not be published in peer-reviewed academic publication outlets~\citeReferencesSectionSecond{yasin2020UsingGreyLiterature}.

\begin{figure}[t]
	\begin{center}
  		\includegraphics[width=0.95\textwidth]{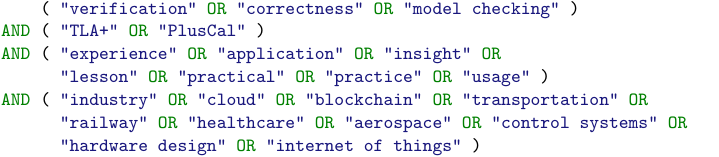}
	\end{center}
	\listingcaption{Google Scholar search string used for the optimistic search of our SLR.}\label{searchstring}
\end{figure}

The first phase involves creating a search string that is specific yet broad enough for comprehensive results.
We targeted the topic and combined it with overlapping synonyms and industry names (see Listing~\ref{searchstring}).
Besides the hypernym \emph{industry}, we additionally included more specific industry names such as \emph{blockchain} or \emph{aerospace} as these domains are known to entail software use cases predestined for formal methods.
This approach aims to optimistically balance the reach and relevance of the search results.
We further limited the search using two inclusion criteria: (1) English language and (2) published within the last 11 years (2013-2023).
After configuring the search settings according to these criteria, we collected a total of 290 papers\footnote{Measurement from last execution on May 28, 2024.}.

\subsection{Reduction}

After the initial paper gathering phase, we reduced the scope in the three consecutive phases, namely deduplication, exclusion, and affinity filter.

\emph{Deduplication:} First, a total of 29 duplicate papers were removed.
This concerned, for example, papers with prior preprints or search results listing a paper once isolated and once in the context of an entire volume such as conference proceedings.

\emph{Rejection:} In a second phase, we rejected papers that did match any of our four exclusion criteria.
Namely, these are (1) no access to paper over university network, (2) not covering a case in industry, (3) neither \TLA nor PlusCal are the selected formal methods, or (4) neither \TLA nor PlusCal appear in abstract, introduction, conclusion or discussion.
A total of 224 papers were rejected in this phase.

\emph{Affinity Filter:} The final reduction phase involved the quality assessment.
However, we denote this phase as \emph{affinity filter} as it concerns the paper's suitability for answering our research questions rather than the paper's overall objective quality.
We quantified a given paper's affinity using four questions, which can either be fully (1 point), partially (0.5 points) or not answered (0 points).
Specifically, we questioned whether the paper addressed: (1) the reason for using formal methods, (2) assumptions for creating the model, (3) the link between model and implementation, and (4) drawbacks from practical experiences.
Papers with $\geq 3.5$ points passed this affinity filter, reducing the set of 37 accepted papers further by 21.
The resulting 16 accepted high-affinity papers are included in this SLR\footnote{See data set on \href{https://doi.org/10.5281/zenodo.13629185}{doi.org/10.5281/zenodo.13629185}.}.
To differentiate these from other references in this paper, we have numbered them accordingly (1-16) and subdivided the bibliography into two sections.

\begin{figure}[t]
	\begin{center}
		\includegraphics[width=0.94\textwidth,trim={0.5cm 0.1cm 0.5cm 0.1cm}]{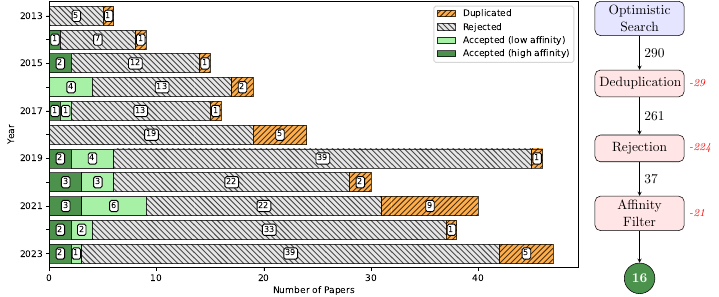}
		\figurecaption{Paper qualifying process, resulting in 16 relevant ones for this SLR.}\label{papercollectionprocess}
	\end{center}
\end{figure}

\section{Results}

The 16 included papers could be linked to some well-known companies or products.
This includes Abaco Autoscaler~\citeReferencesSectionFirst{padhy2022DesigningProvingProperties}, Alibaba Group (Alibaba PolarDB)~\citeReferencesSectionFirst{gu2022CompositionalModelChecking}, Amazon Web Services (including S3 and DynamoDB)~\citeReferencesSectionFirst{newcombe2015HowAmazonWeb}, Audi~\citeReferencesSectionFirst{jakobs2021FollowingWhiteRabbit}, eBay~\citeReferencesSectionFirst{roohitavaf2019LogPlayerFaulttolerantExactlyoncea}, Huawei Cloud (Taurus Distributed Database, now called GaussDB)~\citeReferencesSectionFirst{gao2021FormalVerificationConsensus}, Informal Systems~\citeReferencesSectionFirst{braithwaite2020TendermintLightClient}, Microsoft (Cosmos DB)~\citeReferencesSectionFirst{hackett2023GoingIncidentReport,hackett2023UnderstandingInconsistencyAzure}, MongoDB~\citeReferencesSectionFirst{davis2020EXtremeModellingPractice,schultz2021DesignAnalysisLogless}, PharOS~\citeReferencesSectionFirst{methni2015SpecifyingVerifyingConcurrent} and Thales~\citeReferencesSectionFirst{resch2017UsingTLADevelopment}.
The remainder of this section highlights interesting findings and answers the four research questions stated at the end of Section~\ref{intro}.

\subsection{RQ1: Trend}

The number of publications on \TLA in industrial settings has been growing over the last decade, as shown in Figure~\ref{papercollectionprocess}.
Although only 16 papers qualified for this SLR, the initially considered 290 papers support this claim.
A linear regression analysis of the total unique papers points to statistical significance of the upward trend.
Specifically, the total deduplicated papers (261) exhibit a $S=3.67$ upward slope with $p<0.001$.

This insight of a growing overall trend, however, should be interpreted with caution for two reasons.
First, the adoption of innovative technologies typically follows non-linear evolution curves or hype cycles~\citeReferencesSectionSecond{dedehayir2016HypeCycleModel,rogers2003DiffusionInnovations}.
Given the relatively short time frame considered in this SLR, accurately pinpointing the current stage of this cycle is challenging.
Second, the overall growth trend does not directly address \TLA usage in industrial settings, as posed in this research question.
Nevertheless, the increasing number of unique papers listed in Google Scholar matching the search string is noteworthy, as it suggests ongoing research activity related to the topic.

From the perspective of industry practice only, we record that this growing overall trend resonates in some of the 16 high-affinity papers.
For instance, the emerging trend is carried from within MongoDB~\citeReferencesSectionFirst{davis2020EXtremeModellingPractice}, Huawei~\citeReferencesSectionFirst{gao2021FormalVerificationConsensus}, and Amazon~\citeReferencesSectionFirst{newcombe2015HowAmazonWeb} as they convinced other partners to adopt formal methods.
The Amazon paper by Newcombe et al.~\citeReferencesSectionFirst{newcombe2015HowAmazonWeb} appears particularly impactful, as 7 out of the 16 considered papers (44\%\footnote{All percentages in this paper are rounded to the nearest integer for readability.}) referenced their work.
Gao et al.~\citeReferencesSectionFirst{gao2021FormalVerificationConsensus} explicitly stated it as a key motivator.
This showcases how industry reports can catalyze the adoption of similar practices by other companies.

\begin{tcolorbox}[mybox]
	\emph{Summarized answer to RQ1:}
	There is a significant upward trend in \TLA papers in general, which is likely to continue due to success stories recommending its adoption in industry.
	The report by \mbox{Amazon~\citeReferencesSectionFirst{newcombe2015HowAmazonWeb}} can be seen as a guiding factor in this development.
\end{tcolorbox}

\subsection{RQ2: Application Settings}

We identified that 8 of the papers (50\%) applied pure \TLA while sole PlusCal applications were only recorded in 3 papers (19\%).
The remaining 5 papers (31\%) mention to use both languages in combination.
Papers that utilize PlusCal justify this solely by its intuitive and programmer-friendly syntax.
They appreciate its readability for engineers~\citeReferencesSectionFirst{jakobs2021FollowingWhiteRabbit,resch2017UsingTLADevelopment} without extensive training in formal methods.
Consequently, it is also said to be easier to start with~\citeReferencesSectionFirst{newcombe2014WhyAmazonChose}.
Reasons for using plain \TLA include its records of successful examples in industry~\citeReferencesSectionFirst{gao2021FormalVerificationConsensus} or because it has previously been used within the company~\citeReferencesSectionFirst{schultz2021DesignAnalysisLogless}.

To analyze the domain of application, we classified each of the 16 high-affinity papers based on their most fitting industry affiliation based on an open card sorting approach~\citeReferencesSectionSecond{zimmermann2016CardsortingTextThemes}.
While these industry categories resemble those in our initial search string, they were chosen independently and are unrelated to its design.
The search string was crafted for broad scoping, whereas the industry affiliations of high-affinity papers aim to cluster the \TLA application domains.

We found the majority focuses on cloud applications (63\%), followed by 2 papers for railway and control systems (13\% each), and only 1 paper for blockchain and transportation (6\% each).
This distribution aligns with \TLAno's suitability for distributed systems~\citeReferencesSectionFirst{davis2020EXtremeModellingPractice,gu2022CompositionalModelChecking} and the fact that today's heavy reliance on cloud infrastructure on software-intensive systems makes minimizing downtime crucial~\citeReferencesSectionSecond{buyya2018ManifestoFutureGeneration}.

In addition to industry affiliation, we examined the stage in the system lifecycle where \TLA was predominantly applied.
We depicted these stages with three labels and assigned one or more of them to each of the 16 papers.
Namely, these labels are \emph{early design}, \emph{implementation}, and \emph{debugging}.
By counting the total number of label occurrences, we conclude that \TLA was mostly used during early design (81\%), followed by debugging (44\%), and implementation (38\%).
From these 16 papers, 9 were labelled only with one purpose (7 for early design and 2 for debugging) while the remaining 7 exhibited a combination of purposes.

Keeping a model in sync with an actual implementation is costly.
Therefore, we specifically checked if a paper highlights the link between model and implementation during the affinity filtering phase.
Notably, 3 papers (19\%) discussed automated synchronization between \TLA specifications and productive system implementation.
Salierno et al.~\citeReferencesSectionFirst{salierno2020SpecificationVerificationRailway} proposed a translation algorithm to generate relay logic\footnote{Referring to control systems using electrical relays to perform logical operations.} from the model to verify interlocking components.
Methni et al.~\citeReferencesSectionFirst{methni2015SpecifyingVerifyingConcurrent} introduced \texttt{C2TLA+}, a tool that automatically derives \TLA specification from given \texttt{C} code.
Similarly, Resch and Paulitsch~\citeReferencesSectionFirst{resch2017UsingTLADevelopment} translated \texttt{C} code to PlusCal using scripts.

\begin{tcolorbox}[mybox]
	\emph{Summarized answer to RQ2:}
		\TLA is mostly applied directly or in combination with PlusCal.
		Its predominant application area resides in the cloud industry.
		It is mainly used during the early system design stage, followed by debugging purposes.
		Despite its importance, efforts to synchronize models with implementations have been only marginally addressed.
\end{tcolorbox}

\subsection{RQ3: Fulfillment of Promised Benefits}

All papers highlight several benefits of using \TLAno.
Primarily, it effectively helped to identify subtle system errors or bugs that are difficult to reproduce with conventional testing methods.
Examples include fixing stack overflow errors ~\citeReferencesSectionFirst{davis2020EXtremeModellingPractice}, race-conditions~\citeReferencesSectionFirst{padhy2022DesigningProvingProperties}, or deadlocks~\citeReferencesSectionFirst{roohitavaf2019LogPlayerFaulttolerantExactlyoncea}.
Jakobs et al.~\citeReferencesSectionFirst{jakobs2021FollowingWhiteRabbit} emphasize that, unlike testing, which can only be performed after development, formal verification can be applied from the beginning.

The benefits of improved system design and a better overall understanding of system behavior were equally prominent.
Examples include non-trivial system executions~\citeReferencesSectionFirst{braithwaite2020TendermintLightClient}, considering all possible system states in aggregate~\citeReferencesSectionFirst{hackett2023GoingIncidentReport}, and detecting anomalous but correct behaviors not mentioned in the system's documentation~\citeReferencesSectionFirst{hackett2023UnderstandingInconsistencyAzure}.
Furthermore, Salierno et al.~\citeReferencesSectionFirst{salierno2020SpecificationVerificationRailway} benefited from \TLAno's hardware modelling capabilities and Newcombe et al.~\citeReferencesSectionFirst{newcombe2015HowAmazonWeb} could confidently perform aggressive system optimizations.

\begin{tcolorbox}[mybox]
	\emph{Summarized answer to RQ3:}
	\TLA is reported to successfully identify bugs, contribute to better system design, and overall system understanding -- therewith validating its potential.
\end{tcolorbox}

\clearpage

\subsection{RQ4: Identified Challenges}

On the drawback side, two major aspects were mentioned.
One is the steep learning curve of formal methods, as software engineers are usually unfamiliar with them~\citeReferencesSectionFirst{gao2021FormalVerificationConsensus,resch2017UsingTLADevelopment}.
Although PlusCal was praised in the papers considered for mitigating this, modeling existing systems remains an effortful task that threatens scalability~\citeReferencesSectionFirst{jakobs2021FollowingWhiteRabbit}.

Another challenge is choosing the right level of abstraction~\citeReferencesSectionFirst{gao2021FormalVerificationConsensus} and necessary assumptions~\citeReferencesSectionFirst{roohitavaf2019LogPlayerFaulttolerantExactlyoncea} for the formal model.
Overly detailed models are costly and prone to state space explosion~\citeReferencesSectionSecond{clarke2012ModelCheckingState}, while overly simple models reduce the added value of applying formal methods.
Note that compositional model checking was reported as a technique to help to mitigate state space explosion~\citeReferencesSectionFirst{gu2022CompositionalModelChecking}.
Sabraoui et al.~\citeReferencesSectionFirst{sabraoui2019ModelingMachineCheckingBumpintheWire} also followed such a component-based architecture but later combined them to test the model as a whole.

\begin{tcolorbox}[mybox]
	\emph{Summarized answer to RQ4:}
	\TLAno's steep learning curve, compared to common imperative languages, demands notable effort, especially when applied retrospectively.
	Besides this, selecting the right level of model abstraction remains challenging.
\end{tcolorbox}

\section{Threats to Validity}

As a threat to validity, we acknowledge that other industrial uses of \TLA may exist but are not publicly reported due to intellectual property (IP) issues.
As such, we only considered papers indexed by Google Scholar.
We noticed that using this search engine for the SLR presented challenges, such as slightly varying search results and query rate limits.

In addressing RQ1, we concluded that there is a significant upward trend in the use of \TLAno.
	However, this conclusion is based on the deduplicated number of papers (261) rather than the 16 high-affinity papers, due to the latter's poor sample size.
	The larger set, however, may include irrelevant or nonsensical papers, which potentially skews our trend analysis and leading to misleading conclusions.

	For RQ2, RQ3, and RQ4, our conclusions are derived from manually scanned arguments concerning benefits, challenges, or the affiliations regarding industry and deployment stages.
	As with any manual process, there is a risk of incompleteness or subjective interpretation.
	As mentioned earlier, we mitigated this risk by having two researchers independently conduct the SLR to ensure consistency and reduce bias.
	However, this risk cannot be completely eliminated.

	Finally, it is important to acknowledge that there are likely very few papers that exclusively document the limitations that make \TLA impractical for industry use.
	As a result, the identified high-affinity papers may be positively biased towards the usefulness of \TLA as they primarily showcase success stories.

\section{Conclusion and Future Work}

With this SLR we analyzed the surface of \TLAno's industry adoption using our RQs.
We conclude that \TLA usage has surged since 2015, especially in the cloud industry and is mainly applied during early system design and debugging.
It effectively uncovers deep bugs, enhances system design, and improves overall understanding.
However, its adoption is hindered by a steep learning curve and the challenge of selecting appropriate model abstractions, despite PlusCal's help in easing this difficulty.

With this contribution we shed light on the actual use of formal methods in industry and its consequences, focusing on \TLA and the last decade.
We believe the resulting curated list of relevant papers and our findings serve as a foundation for two stakeholders.
First, interested practitioners that have more specific questions in favor of a potential adoption can use it to bootstrap their consultation.
Second, the identified reported challenges from adopting \TLA serve academics for directing future optimization efforts effectively.

Based on the curated list of papers, future work could more broadly address practical challenges and produce valuable lessons learned regarding introducing formal methods in industrial settings.
Case studies, structured interviews of experts and other empirical studies with industry software engineers and formal method researchers could explore their experiences with system design issues and hard-to-reproduce bugs when using formal methods.
Such investigations would improve understanding of their needs and promote \TLA adoption in industrial software.

Regarding the systematic literature review performed in this paper, future work can mitigate limitations and address threats to validity.
Naturally, extending the time range beyond 10 years, to other search engines and beyond scientific works may provide further insights.
Findings themselves as presented in this paper can be further confirmed through other empirical methods, such as expert surveys, interviews or through informal company documents.
Generality of findings can be also increased by employing a less restrictive affinity filter.
Finally, we addressed \TLA and its use in practice.
Analyzing such trends in the industrial application of formal methods in general can help assess (and ensure) that their benefits resonate with practitioners and can be valuable for the broader scientific community.

\begin{credits}
	\pdfbookmark[1]{\ackname}{ack} 
	\subsubsection*{\ackname}
		This work has been supported by the Swiss National Science Foundation (SNSF) within project ``\texttt{RUNVERSPACE}: Runtime Verification for Space Software Architectures'', grant no. 220875.
\end{credits}

\clearpage

\bibliographystyleReferencesSectionFirst{splncs04}
\bibliographyReferencesSectionFirst{references.bib}

\bibliographystyleReferencesSectionSecond{splncs04}
\bibliographyReferencesSectionSecond{references.bib} 


\begin{thebibliography}{10}
\providecommand{\url}[1]{\texttt{#1}}
\providecommand{\urlprefix}{URL }
\providecommand{\doi}[1]{https://doi.org/#1}

\bibitem{braithwaite2020TendermintLightClient}
Braithwaite, S., Buchman, E., Khoffi, I., Konnov, I., Milosevic, Z., Ruetschi, R., Widder, J.: A {Tendermint} {Light} {Client} (Oct 2020). \doi{10.48550/arXiv.2010.07031}

\bibitem{davis2020EXtremeModellingPractice}
Davis, A.J.J., Hirschhorn, M., Schvimer, J.: {eXtreme} {Modelling} in {Practice}. Proceedings of the VLDB Endowment  \textbf{13}(9),  1346--1358 (May 2020). \doi{10.14778/3397230.3397233}

\bibitem{gao2021FormalVerificationConsensus}
Gao, S., Zhan, B., Liu, D., Sun, X., Zhi, Y., Jansen, D.N., Zhang, L.: Formal {Verification} of {Consensus} in the {Taurus} {Distributed} {Database}. In: Huisman, M., Păsăreanu, C., Zhan, N. (eds.) Formal {Methods}. pp. 741--751. Springer International Publishing, Cham (2021). \doi{10.1007/978-3-030-90870-6_42}

\bibitem{gu2022CompositionalModelChecking}
Gu, X., Cao, W., Zhu, Y., Song, X., Huang, Y., Ma, X.: Compositional {Model} {Checking} of {Consensus} {Protocols} via {Interaction}-{Preserving} {Abstraction}. In: 2022 41st {International} {Symposium} on {Reliable} {Distributed} {Systems} ({SRDS}). pp. 82--93 (Sep 2022). \doi{10.1109/SRDS55811.2022.00018}

\bibitem{hackett2023GoingIncidentReport}
Hackett, F., Rowe, J., Kuppe, M.A.: Going {Beyond} an {Incident} {Report} with {TLA}$^{\textrm{+}}$ (Jul 2023), \url{https://www.usenix.org/sites/default/files/login_-_going_beyond_an_incident_report_with_tla_.pdf}

\bibitem{hackett2023UnderstandingInconsistencyAzure}
Hackett, F., Rowe, J., Kuppe, M.A.: Understanding {Inconsistency} in {Azure} {Cosmos} {DB} with {TLA}+. In: 2023 {IEEE}/{ACM} 45th {International} {Conference} on {Software} {Engineering}: {Software} {Engineering} in {Practice} ({ICSE}-{SEIP}). pp. 1--12. IEEE (2023). \doi{10.1109/ICSE-SEIP58684.2023.00006}

\bibitem{jakobs2021FollowingWhiteRabbit}
Jakobs, C., Werner, M., Schmidt, K., Hansch, G.: Following the {White} {Rabbit}: {Integrity} {Verification} {Based} on {Risk} {Analysis} {Results}. In: Proceedings of the 5th {ACM} {Computer} {Science} in {Cars} {Symposium}. pp.~1--9. {CSCS} '21, Association for Computing Machinery, New York, NY, USA (Nov 2021). \doi{10.1145/3488904.3493377}

\bibitem{methni2015SpecifyingVerifyingConcurrent}
Methni, A., Lemerre, M., Ben~Hedia, B., Haddad, S., Barkaoui, K.: Specifying and {Verifying} {Concurrent} {C} {Programs} with {TLA}+. In: Artho, C., Ölveczky, P.C. (eds.) Formal {Techniques} for {Safety}-{Critical} {Systems}. pp. 206--222. Springer International Publishing, Cham (2015). \doi{10.1007/978-3-319-17581-2_14}

\bibitem{newcombe2014WhyAmazonChose}
Newcombe, C.: Why {Amazon} {Chose} {TLA}$^{\textrm{+}}$. In: Ait~Ameur, Y., Schewe, K.D. (eds.) Abstract {State} {Machines}, {Alloy}, {B}, {TLA}, {VDM}, and {Z}. pp. 25--39. Springer, Berlin, Heidelberg (2014). \doi{10.1007/978-3-662-43652-3_3}

\bibitem{newcombe2015HowAmazonWeb}
Newcombe, C., Rath, T., Zhang, F., Munteanu, B., Brooker, M., Deardeuff, M.: How {Amazon} web services uses formal methods. Communications of the ACM  \textbf{58}(4),  66--73 (Mar 2015). \doi{10.1145/2699417}

\bibitem{padhy2022DesigningProvingProperties}
Padhy, S., Stubbs, J.: Designing and {Proving} {Properties} of the {Abaco} {Autoscaler} {Using} {TLA}+. In: Bloem, R., Dimitrova, R., Fan, C., Sharygina, N. (eds.) Software {Verification}. pp. 86--103. Springer International Publishing, Cham (2022). \doi{10.1007/978-3-030-95561-8_6}

\bibitem{resch2017UsingTLADevelopment}
Resch, S., Paulitsch, M.: Using {TLA}+ in the {Development} of a {Safety}-{Critical} {Fault}-{Tolerant} {Middleware}. In: 2017 {IEEE} {International} {Symposium} on {Software} {Reliability} {Engineering} {Workshops} ({ISSREW}). pp. 146--152 (Oct 2017). \doi{10.1109/ISSREW.2017.43}

\bibitem{roohitavaf2019LogPlayerFaulttolerantExactlyoncea}
Roohitavaf, M., Ren, K., Zhang, G., Ben-romdhane, S.: {LogPlayer}: {Fault}-tolerant {Exactly}-once {Delivery} using {gRPC} {Asynchronous} {Streaming} (Nov 2019). \doi{10.48550/arXiv.1911.11286}

\bibitem{sabraoui2019ModelingMachineCheckingBumpintheWire}
Sabraoui, M., Hieb, J., Lauf, A., Graham, J.: Modeling and {Machine}-{Checking} {Bump}-in-the-{Wire} {Security} for {Industrial} {Control} {Systems}. In: Staggs, J., Shenoi, S. (eds.) Critical {Infrastructure} {Protection} {XIII}. pp. 271--288. Springer International Publishing, Cham (2019). \doi{10.1007/978-3-030-34647-8_14}

\bibitem{salierno2020SpecificationVerificationRailway}
Salierno, G., Morvillo, S., Leonardi, L., Cabri, G.: Specification and verification of railway safety-critical systems using {TLA}$^{\textrm{+}}$: {A} {Case} {Study}. In: 2020 {IEEE} 29th {International} {Conference} on {Enabling} {Technologies}: {Infrastructure} for {Collaborative} {Enterprises} ({WETICE}). pp. 207--212 (Sep 2020). \doi{10.1109/WETICE49692.2020.00048}

\bibitem{schultz2021DesignAnalysisLogless}
Schultz, W., Zhou, S., Dardik, I., Tripakis, S.: Design and {Analysis} of a {Logless} {Dynamic} {Reconfiguration} {Protocol} (Nov 2021). \doi{10.48550/arXiv.2102.11960}

\end{thebibliography}


\begin{thebibliography}{10}
\providecommand{\url}[1]{\texttt{#1}}
\providecommand{\urlprefix}{URL }
\providecommand{\doi}[1]{https://doi.org/#1}

\bibitem{buyya2018ManifestoFutureGeneration}
Buyya, R., Srirama, S.N., Casale, G., Calheiros, R., Simmhan, Y., Varghese, B., Gelenbe, E., Javadi, B., Vaquero, L.M., Netto, M.A.S., Toosi, A.N., Rodriguez, M.A., Llorente, I.M., Vimercati, S.D.C.D., Samarati, P., Milojicic, D., Varela, C., Bahsoon, R., Assuncao, M.D.D., Rana, O., Zhou, W., Jin, H., Gentzsch, W., Zomaya, A.Y., Shen, H.: A {Manifesto} for {Future} {Generation} {Cloud} {Computing}: {Research} {Directions} for the {Next} {Decade}. ACM Computing Surveys  \textbf{51}(5),  105:1--105:38 (Nov 2018). \doi{10.1145/3241737}

\bibitem{carrera-rivera2022HowtoConductSystematic}
Carrera-Rivera, A., Ochoa, W., Larrinaga, F., Lasa, G.: How-to conduct a systematic literature review: {A} quick guide for computer science research. MethodsX  \textbf{9},  101895 (Jan 2022). \doi{10.1016/j.mex.2022.101895}

\bibitem{clarke2012ModelCheckingState}
Clarke, E.M., Klieber, W., Nováček, M., Zuliani, P.: Model {Checking} and the {State} {Explosion} {Problem}. In: Meyer, B., Nordio, M. (eds.) Tools for {Practical} {Software} {Verification}: {LASER}, {International} {Summer} {School} 2011, {Elba} {Island}, {Italy}, {Revised} {Tutorial} {Lectures}, pp. 1--30. Springer, Berlin, Heidelberg (2012). \doi{10.1007/978-3-642-35746-6_1}

\bibitem{dedehayir2016HypeCycleModel}
Dedehayir, O., Steinert, M.: The hype cycle model: {A} review and future directions. Technological Forecasting and Social Change  \textbf{108},  28--41 (Jul 2016). \doi{10.1016/j.techfore.2016.04.005}

\bibitem{kitchenham2007GuidelinesPerformingSystematic}
Kitchenham, B.A., Charters, S.: Guidelines for performing systematic literature reviews in software engineering. Tech. Rep. EBSE 2007-001, Keele University and Durham University Joint Report (07 2007), \url{https://www.elsevier.com/__data/promis_misc/525444systematicreviewsguide.pdf}

\bibitem{lamport2002specifying}
Lamport, L.: Specifying {Systems}: {The} {TLA}+ {Language} and {Tools} for {Hardware} and {Software} {Engineers}. Addison-Wesley (Jun 2002), \url{https://www.microsoft.com/en-us/research/uploads/prod/2018/05/book-02-08-08.pdf}

\bibitem{lamport2009PlusCalAlgorithmLanguage}
Lamport, L.: The {PlusCal} {Algorithm} {Language}. In: Leucker, M., Morgan, C. (eds.) Theoretical {Aspects} of {Computing} - {ICTAC} 2009. pp. 36--60. Springer, Berlin, Heidelberg (2009). \doi{10.1007/978-3-642-03466-4_2}

\bibitem{petersen2015GuidelinesConductingSystematic}
Petersen, K., Vakkalanka, S., Kuzniarz, L.: Guidelines for conducting systematic mapping studies in software engineering: {An} update. Information and Software Technology  \textbf{64},  1--18 (Aug 2015). \doi{10.1016/j.infsof.2015.03.007}

\bibitem{pnueli1977TemporalLogicPrograms}
Pnueli, A.: The temporal logic of programs. In: 18th {Annual} {Symposium} on {Foundations} of {Computer} {Science} (sfcs 1977). pp. 46--57 (Oct 1977). \doi{10.1109/SFCS.1977.32}

\bibitem{reid2020MakingFormalMethods}
Reid, A., Church, L., Flur, S., de~Haas, S., Johnson, M., Laurie, B.: Towards making formal methods normal: meeting developers where they are (Oct 2020). \doi{10.48550/arXiv.2010.16345}

\bibitem{rogers2003DiffusionInnovations}
Rogers, E.M.: Diffusion of {Innovations}. Social science, Free Press, New York London Toronto Sydney, 5 edn. (2003), {I}SBN: 978-0-7432-2209-9

\bibitem{yasin2020UsingGreyLiterature}
Yasin, A., Fatima, R., Wen, L., Afzal, W., Azhar, M., Torkar, R.: On {Using} {Grey} {Literature} and {Google} {Scholar} in {Systematic} {Literature} {Reviews} in {Software} {Engineering}. IEEE Access  \textbf{8},  36226--36243 (2020). \doi{10.1109/ACCESS.2020.2971712}

\bibitem{zimmermann2016CardsortingTextThemes}
Zimmermann, T.: Card-sorting: {From} text to themes. In: Menzies, T., Williams, L., Zimmermann, T. (eds.) Perspectives on {Data} {Science} for {Software} {Engineering}, pp. 137--141. Morgan Kaufmann, Boston (Jan 2016). \doi{10.1016/B978-0-12-804206-9.00027-1}

\end{thebibliography}
\end{document}